\documentclass[submission,copyright,creativecommons]{eptcs}
\providecommand{\event}{CREST 2016} 
\usepackage{breakurl}             
\usepackage{epsfig}
\usepackage{graphicx}
\usepackage{wrapfig}
\usepackage{xspace}

\newtheorem{theorem}{Theorem}[section]

\newtheorem{definition}[theorem]{Definition}

\newenvironment{RETHM}[2]{\it \trivlist \item[\hskip \labelsep{\bf #1 \ref{#2}}]}{\endtrivlist}
\newcommand{\rethm}[1]{\begin{RETHM}{Theorem}{#1}}
\newcommand{\repro}[1]{\begin{RETHM}{Proposition}{#1}}
\newcommand{\relem}[1]{\begin{RETHM}{Lemma}{#1}}
\newcommand{\recor}[1]{\begin{RETHM}{Corollary}{#1}}

\newcommand{\erethm}{\end{RETHM}}
\newcommand{\erepro}{\end{RETHM}}
\newcommand{\erelem}{\end{RETHM}}
\newcommand{\erecor}{\end{RETHM}}

\def\squarebox#1{\hbox to #1{\hfill\vbox to #1{\vfill}}}
\newcommand{\qed}{\hspace*{\fill}
            \vbox{\hrule\hbox{\vrule\squarebox{.667em}\vrule}\hrule}\smallskip}
\newenvironment{proof}{\begin{trivlist}
\item[\hspace{\labelsep}{\bf\noindent Proof: }]
}{\qed\end{trivlist}}

\newtheorem{xmpl}[theorem]{Example}

\newtheorem{rmark}[theorem]{Remark}

\newcommand{\U}{{\bf U}}

\newcommand{\cS}{{\bf S}}

\newcommand{\zug}[1]{\langle #1  \rangle}

\newcommand{\stam}[1]{}

\newcommand{\bd}{\begin{definition}}
\newcommand{\ed}{\end{definition}}
\newcommand{\be}{\begin{enumerate}}
\newcommand{\bi}{\begin{itemize}}
\newcommand{\ee}{\end{enumerate}}
\newcommand{\ei}{\end{itemize}}
\newcommand{\T}{\mbox{\sc \bf true}\xspace}
\newcommand{\F}{\mbox{\sc \bf false}\xspace}

\newcommand{\dKw}{\tilde{K}_{w,q}}

\newcommand{\cF}{{\bf F}}
\newcommand{\V}{{\bf V}}
\newcommand{\R}{{\bf R}}

\newcommand{\G}{{\bf G}}

\renewcommand{\phi}{\varphi}

\newcommand{\dr}{\mbox{{\em dr}}}
\newcommand{\db}{\mbox{{\em db}}}
\newcommand{\K}{{\it K}}

\begin{document}

\title{Causality and Responsibility \\ for Formal Verification and Beyond}
\author{Hana Chockler
\institute{Department of Informatics \\
King's College London}
\email{hana.chockler@kcl.ac.uk}
}
\def\titlerunning{Causality for Formal Verification}
\def\authorrunning{Hana Chockler}

\maketitle

\begin{abstract}
The theory of actual causality, defined by Halpern and Pearl, and its quantitative measure -- the degree
of responsibility -- was shown to be 
extremely useful in various areas of computer science due to a good match between the results it produces and our intuition. In this paper, I describe the applications of causality to formal verification, namely, explanation of counter-examples, refinement of coverage metrics,
and symbolic trajectory evaluation. 
I also briefly discuss recent applications of causality to legal reasoning.
\end{abstract}

\section{Introduction}

The definition of causality given by Halpern and Pearl \cite{HP01b}, like other
definitions of causality in the philosophy literature going back to Hume
\cite{Hum39}, is based on {\em counterfactual dependence}.  
Essentially,
event $A$ is a cause of event $B$ if, had $A$ not happened (this is the
counterfactual condition, since $A$ did in fact happen) then $B$ would
not have happened.  Unfortunately, this definition does not capture all
the subtleties involved with causality.  
For example, suppose 
that
Suzy and Billy both pick up rocks
and throw them at a bottle ( the example is due to Hall \cite{Hall98}).
Suzy's rock gets there first, shattering the
bottle.  Since both throws are perfectly accurate, Billy's would have
shattered the bottle had it not been preempted by Suzy's throw.
(This story is taken from \cite{Hall98}.)  Thus, according to 
the counterfactual condition, 
Suzy's throw is not a 
cause for shaterring the bottle.
This problem is dealt with in \cite{HP01b}  
by, roughly speaking, 
taking
$A$ to be a cause of $B$ if $B$ counterfactually depends on $A$ under
some contingency.  For example, Suzy's throw is 
a cause of the bottle
shattering because the bottle shattering counterfactually depends on
Suzy's throw, under the contingency that Billy doesn't throw.
It may seem that this solves one problem only to create another.
While this allows Suzy's throw to be a cause of the bottle shattering,
it also seems to allow Billy's throw to be a cause too, which seems counter-intuitive
to most people. As is shown in
\cite{HP01}, it is possible to build a more sophisticated model that expresses the subtlety
of pre-emption in this case, using auxiliary variables to express the order in which
the rocks hit the bottle and preventing Billy's throw from being a cause of the bottle
shattering. One moral of this example
is that, according to the \cite{HP01} definitions, whether or not $A$ is
a cause of $B$ depends in part on the model used.  

Halpern and Pearl's definition of causality, while extending and refining the counter-factual
definition, still treats causality  as an all-or-nothing concept.
That is, $A$ is either a cause of $B$ or it is not.  The concept of responsibility,
introduced in~\cite{CH04}, presents a way to quantify causality and hence the
ability to measure the degree of influence (aka ``the degree of responsibility'') of different
causes on the outcome. For
example, suppose that Mr.~B wins an election against Mr.~G. If the vote
has been  11--0, it is clear that each of the people who voted for Mr.~B
is a cause of him winning, but the degree of responsibility of each voter for
Mr.~B is lower than in a vote 6--5. 

Thinking in terms of causality and responsibility was shown to be beneficial for a
seemingly unrelated area of research, namely formal verification (model checking)
of computerised systems. 
In {\em model checking}, we verify the
correctness of a  finite-state system with respect to a desired
behavior by checking whether a labeled state-transition graph that
models the system satisfies a specification of this behavior
\cite{CGP99}. If the answer to the correctness query is negative, the tool provides
a counterexample to the satisfaction of the specification in the system.
These counterexamples are used for debugging the system, as
they demonstrate an example of an erroneous behaviour~\cite{CGMZ95}.
As counterexamples can be very long and complex, there is a need for an additional
tool explaining ``what went wrong'', that is, pinpointing the \emph{causes} of
an error on the counterexample.
As I describe in more details in the following sections, the causal analysis of 
counterexamples, described in~\cite{BBCOT12}, is an integral part of an industrial
hardware verification platform RuleBase of IBM~\cite{RBurl}. 

On the other hand, if the answer is positive, the tools usually perform some type of a sanity check,
to verify that the positive result was not caused by an error or underspecification (see ~\cite{Kup06} for a survey). \emph{Vacuity} check, which is the most common sanity check and is a standard in industrial
model checking tools, was first defined by Beer et al.~\cite{BBER01} as a situation, where the property
passes in a ``non-interesting way'', that is, a part of a property does not affect the model checking
procedure in the system. Beer et al. state that vacuity was a serious problem in verification of hardware
designs at IBM: ``our experience has shown that typically 20\% of specifications 
pass vacuously during the first formal-verification runs of a new hardware
design, and that vacuous passes always point to a real problem in either the design
or its specification or environment'' \cite{BBER01}. The general vacuity problem was formalised by 
Kupferman and Vardi, who defined a vacuous pass as a pass, where some subformula of the original
property can be replaced by its $\bot$ value without affecting the satisfaction of the property
in the system \cite{KV03a}, and there is a plethora of subsequent papers and definitions addressing
different aspects and nuances of vacuous satisfaction. 

Note that vacuity, viewed from the point of view of
causal analysis, is \emph{counterfactual causality}. Indeed, a property passes non-vacuously if
each its subformula, if replaced by $\bot$, causes falsification of the property in the system.
Due to the nature of specifications as more general than implementations, it is quite natural
to expect that the specification allows some degree of freedom in how it is going to be satisfied.
Consider, for example, a specification $\G(p \vee q)$, meaning
that either $p$ or $q$ should hold in each state. If in the system under verification both $p$ and $q$
are \T in all states, the standard vacuity check will alert the verification engineer to vacuity in $p$
and in $q$. In contrast, introducing a contingency where $p$ is set to \F causes the result of model
checking to counterfactually depend on $q$ (and similarly for $q$ and $p$), indicating that both $p$
and $q$ play some role in the satisfaction of the specification in the system.

\emph{Coverage} check is a concept ``borrowed'' from testing and simulation-based verification,
where various coverage metrics are traditionally used as heuristic measures of exhaustiveness of the
verification procedure \cite{TK01}. In model checking, the suitable coverage concept is
\emph{mutation coverage}, introduced by Hoskote et al.~\cite{HKHZ99}, and formalized in
\cite{CKV01,CKKV01,CK02a,CKV03}. In this definition, an element of the system under verification is
considered \emph{covered} by the specification if changing (\emph{mutating}) this element 
falsifies the specification in the system. Note, again, that this definition is, essentially, a counter-factual
causality. As a motivating example to the necessity to finer-grained analysis consider the property
$\cF req$, meaning in every computation there is at least one request. Now consider a system
in which requests are sent frequently, resulting in several requests on each of the computational paths.
All these requests will be considered not covered by the mutation coverage metric, as changing each
one of them separately to \F does not falsify the property; and yet, each of these requests plays
a role in the satisfaction of the specification (or, using the notions from causality, for each request
there exists a contingency that creates a counterfactual dependence between this request and
the satisfaction of the specification) \cite{CHK08}. 

While causality allows to extend the concepts of vacuity and coverage to include elements that
\emph{affect} the satisfaction of the specification in the system in some way, it is still
an all-or-nothing concept. Harnessing the notion of responsibility to measure the
influence of different elements on the success or failure of the model checking process
introduces the quantification aspect, providing a finer-grained analysis. While, as I
discuss in more detail below, the full-blown responsibility computation is intractable for all but
very small systems, introducing a \emph{threshold} on the value of responsibility, in order
to detect only the most influential causes, reduces the complexity and makes the computation
manageable for real systems  \cite{CHK08}. 

 The quantification provided by the notion of responsibility and the distinction between
influential and non-influential causes have been applied to the symbolic trajectory evaluation, where
ordering the  causes by their degree of responsibility was demonstrated to be a good heuristic
for instantiating 
a minimal number of variables that is sufficient to determine the output value of the circuit~\cite{CGY08}.

In the next sections I provide a brief overview of the relevant concepts and describe the applications
of causality to formal verification in more detail.

\section{Causality and Responsibility -- Definitions}\label{sec:definitions}
\label{sec:prelim}

In this section, I briefly review the details of 
Halpern and Pearl's definition (HP) of causal models and causality  \cite{HP01b} and the definitions of
responsibility and blame \cite{CH04} in causal models. 

\subsection{Causal models}

A {\em signature\/} is a tuple $\cS = \zug{\U,\V,\R}$, 
where $\U$ is a finite set
of {\em exogenous\/} variables, $\V$ is a finite
set of {\em endogenous\/}
variables,  
and $\R$ associates with every variable  
$Y \in \U \cup \V$ a finite
nonempty set $\R(Y)$ of possible values for $Y$.
Intuitively, the  exogenous variables are ones whose values are
determined by factors outside the model, while the endogenous variables
are ones whose values are ultimately determined by the exogenous
variables.
A (recursive) {\em causal model\/} over signature $\cS$ is a tuple
$M = \zug{\cS,\cF}$, where $\cF$ associates with every endogenous variable
$X \in \V$ a function $F_X$ such that 
$F_X: (\times_{U \in \U} \R(U) \times (\times_{Y \in \V \setminus \{ X \}}
\R(Y))) \rightarrow \R(X)$, and functions have no circular dependency.
That is, $F_X$ describes how the value of the
endogenous variable $X$ is determined by
the values of all other variables in $\U \cup \V$. 
If all variables have only two values, 
we say that $M$ is a 
{\em binary causal model}.

We can describe (some salient features of) a causal model $M$ using a
{\em causal network},
which is a  graph with nodes corresponding to the variables in $M$ and
edges corresponding to the dependence between variables. 
We focus our attention
on \emph{recursive models} -- those whose associated causal network
is a directed acyclic graph.

A causal network
is a graph
with nodes corresponding to the random variables in $\V$ and an edge
from a node labeled $X$ to one labeled $Y$ if $F_Y$ depends on the value
of $X$.
Intuitively, variables can have a causal effect only on their
descendants in the causal network; if $Y$ is not a descendant of $X$,
then a change in the value of $X$ has no affect on the value of $Y$.
A setting $\vec{u}$ for the variables in $\U$ is called a {\em context}.
It should be clear that if $M$ is a recursive causal model,
then there is always a
unique solution to the equations in $M$, given a context.

Given a causal model $M = (\cS,\cF)$, a (possibly
empty)  vector
$\vec{X}$ of variables in $\V$, and a vector $\vec{x}$ 
of values for the variables in
$\vec{X}$, a new causal model,
denoted $M_{\vec{X} \gets \vec{x}}$, is defined as
identical to $M$, except that the
equation for the variables $\vec{X}$ in $\cF$ is replaced by $\vec{X} =
\vec{x}$. 
Intuitively, this is the causal model that results when the variables in
$\vec{X}$ are set to $\vec{x}$ by some external action
that affects only the variables in $\vec{X}$
(and overrides the effects of the causal equations).

A causal formula $\phi$ -- a Boolean combination of
events capturing the value of variables in the model -- is \T or \F in a causal model, given a
\emph{context}.
We write $(M,\vec{u}) \models \phi$ if
$\phi$ is \T in the
causal model $M$, given the context $\vec{u}$.
$(M,\vec{u}) \models  [\vec{Y} \gets \vec{y}](X = x)$ if 
the variable $X$ has value $x$ 
in the unique solution
to the equations in $M_{\vec{Y} \gets \vec{y}}$ in context $\vec{u}$. 
We now review the HP definition of causality..
\begin{definition}[Cause \cite{HP01b}]\label{def-cause}
$\vec{X} = \vec{x}$ is a {\em cause\/} of $\varphi$ in
$(M,\vec{u})$ if the following three conditions hold: 
\begin{description}
\item[AC1.] $(M,\vec{u}) \models (\vec{X} = \vec{x}) \wedge \varphi$. 
\item[AC2.] There exist a partition $(\vec{Z},\vec{W})$ of $\V$ with 
$\vec{X} \subseteq \vec{Z}$ and some setting 
$(\vec{x}',\vec{w})$ of the
variables in $(\vec{X},\vec{W})$ such that if $(M,\vec{u}) \models Z = z^*$
for $Z \in \vec{Z}$, then
\be
\item[(a)] $(M,\vec{u}) \models [ \vec{X} \leftarrow \vec{x}',
\vec{W} \leftarrow \vec{w}]\neg{\varphi}$. 
\item[(b)] $(M,\vec{u}) \models [ \vec{X} \leftarrow \vec{x},
\vec{W}' \leftarrow \vec{w}, \vec{Z}' \leftarrow \vec{z}^*]\varphi$ for
all subsets $\vec{Z}'$ of $\vec{Z} \setminus \vec{X}$ and all subsets 
$\vec{W}'$ of $\vec{W}$. 
The tuple $(\vec{W}, \vec{w}, \vec{x}')$ is said to be a
\emph{witness} to the fact that $\vec{X} = \vec{x}$ is a cause of $\phi$.
\ee
\item[AC3.] 
$(\vec{X} = \vec{x})$ is minimal; no subset of
$\vec{X}$ satisfies AC2.
\end{description}
\end{definition}

Essentially, Definition~\ref{def-cause} extends the counterfactual definition of 
causality by considering contingencies $\vec{W}$ -- changes in the current context that by
themselves do not change the value of $\varphi$, but create a counterfactual
dependence between the value of a $\vec{X}$ and the value of $\varphi$. The
variables in $\vec{Z}$ should be thought of as
describing the ``active causal process'' from $X$ to $\phi$, and are needed in order
to make sure that, while introducing contingencies, we preserve the causal process from
$\vec{X}$ to $\varphi$. The minimality requirement AC3 is needed in order to avoid
adding irrelevant variables to $\vec{X}$.

We note that Halpern recently updated the definition of
causality, changing the concept to focus on the variables that are \emph{frozen} in their
original values, rather than considering contingencies \cite{Hal15b}. Since
the existing work on the applications of causality to formal verification uses the previous definition,
we continue using it in this paper.
 
\vspace{-0.3cm}
\subsection{Responsibility and Blame}\label{sec-def-resp}

Causality is a ``0--1'' concept; $\vec{X} = \vec{x}$ is either a cause of
$\phi$ or it is not.  
Now consider two voting scenarios: in the first, Mr.~G
beats Mr.~B by a vote of 11--0.  In the second, Mr.~G beats Mr.~B by a
vote of 6--5.  According to the HP
definition, all the people who voted for Mr. G are causes of him
winning.  While this does not seem so unreasonable, it does not capture
the intuition that each voter for Mr. G is more critical to the
victory in the case of the 6--5 vote than in the case of the 11--0 vote.
The notion of \emph{degree of responsibility}, introduced by Chockler and
Halpern \cite{CH04}, extends the notion of causality to capture the differences
in the degree of criticality of causes.
%
In the case of the 6--5 vote, no changes have to be made to make each
voter for Mr.~G critical for Mr.~G's victory; if he had not voted for
Mr.~G, Mr.~G would not have won.  Thus, each voter has degree of
responsibility 1 (i.e., $k=0$).  On the other hand, in the case of the
11--0 vote, for a particular voter to be critical, five other voters
have to switch their votes; thus, $k=5$, and each voter's degree of
responsibility is $1/6$. 
This notion of degree of responsibility has
been shown to capture (at a qualitative level) the way people allocate
responsibility \cite{LGZ13}. 

\begin{definition}[Degree of Responsibility \cite{CH04}]\label{def-resp}
The {\em degree of responsibility
of $\vec{X}=\vec{x}$ for $\phi$ in 
$(M,\vec{u})$\/}, denoted $\dr((M,\vec{u}), (\vec{X} \gets \vec{x}), \phi)$, is
$0$ if $\vec{X}=\vec{x}$ is 
not a cause of $\phi$ in $(M,\vec{u})$; it is $1/(k+1)$ if
$\vec{X}=\vec{x}$ is  a cause of $\phi$ in $(M,\vec{u})$ according
to Definition~\ref{def-cause} with $\vec{W}$ of size $k$ 
being a smallest
witness to the fact that $\vec{X}=\vec{x}$ is a cause of $\phi$.
\end{definition}

When
determining responsibility, it is assumed that everything relevant about
the facts of the world and how the world works (which we characterize in
terms of  {\em structural equations\/})
is known. 
But this misses out on important
component of determining what Chockler and Halpern call  {\em blame}:~the epistemic
state.  
Formally, the degree of blame, introduced by Chockler and Halpern
is the expected degree of responsibility \cite{CH04}.  
This is perhaps best
understood by considering a firing squad with ten excellent marksmen
(the example is by Tim Williamson).  
Only one of them has live bullets in his rifle; the rest have blanks.  
The marksmen do not know which of them has the live bullets.  The
marksmen shoot at the prisoner and he dies.  The only marksman that is
the cause of the prisoner's death is the one with the live bullets.
That marksman has degree of responsibility $1$ for the death; all the rest
have degree of responsibility $0$.  However, each of the marksmen has
degree of blame $1/10$.

An agent's uncertainty is modelled by a pair $(\K,\Pr)$, where $\K$
is a set of pairs of the form $(M,\vec{u})$, where $M$ is a 
causal model and $\vec{u}$ is a context,
and $\Pr$ is a probability distribution over $\K$. Note that probability
is used here in a rather non-traditional sense, to capture the epistemic state of an
agent, rather than an actual probability over values of variables.
\begin{definition}[Blame \cite{CH04}]\label{def-blame}
The {\em degree of 
blame of setting $\vec{X}$ to $\vec{x}$ for $\phi$ relative to epistemic state
$(\K,\Pr)$\/}, denoted $\db(\K,\Pr,\vec{X} \gets \vec{x}, \phi)$, is defined
as an \emph{expected value} of the degree of responsibility over the probability
space $(\K,\Pr)$.
\end{definition} 

\section{Coverage in the framework of causality}\label{cover-cause}

The following definition of coverage is based on the study of {\em mutant coverage\/}
in simulation-based verification 
~\cite{MLS78,MO91,AB01},
and is the one that is adopted in all (or almost all) papers on coverage metrics in
formal verification today (see, for example, 
\cite{HKHZ99,CKV01,CKKV01,CK02a,CKV03}).
For a Kripke structure $K$, an atomic proposition
$q$, and a state $w$, we denote by $\dKw$ the Kripke structure
obtained from $K$ by flipping the value of $q$ in $w$. 

\begin{definition}[Coverage]\label{cov-def}
Consider a Kripke structure $K$, a specification $\varphi$ that is satisfied in
$K$, and an atomic proposition $q \in AP$. A state $w$ of $K$ is 
{\em $q$-covered by $\varphi$\/} if $\dKw$ does not satisfy $\varphi$.
\end{definition}

It is easy to see that coverage corresponds to the simple 
counterfactual-dependence approach to causality. 
Indeed, a state $w$ of $K$ is 
{\em $q$-covered by $\varphi$\/} if $\varphi$ holds in $K$ and if
$q$ had other value in $w$, then $\varphi$ would not have been true in 
$K$. The following example illustrates the notion of coverage
and shows that the counter-factual approach to coverage misses some
important insights in how the system satisfies the specification.
Let $K$ be a Kripke structure presented with one path, where one request is followed by
three grants in subsequent states, and 
let $\phi = \G(req \rightarrow \cF grant)$ (every request is eventually granted).
It is easy to see that $K$
satisfies $\phi$, but that none of the states are covered with respect to $grant$, 
as flipping the value of $grant$ in one of
them does not falsify $\phi$ in $K$. 
On the other hand, representing the model checking procedure as a causal model with a context
corresponding to the actual values of atomic propositions in states (see \cite{CHK08} for a formal
description of this representation), demonstrates that for each state there exists a contingency
where the result counterfactually depends on the value of \emph{grant} in this state; the contingency is
removing grants in the two other states. Hence, while none of the states is covered with respect to
\emph{grant}, they are all causes of $\phi$ in $K$ with the responsibility $1/3$.

In the example above, and typically in the applications of causality to formal verification, there are no
structural equations over the internal endogenous variables. However, if we want to express the
temporal characteristics of our model -- for example, to say that the first \emph{grant} is important, whereas
subsequent ones are not -- the way to do so is by introducing internal auxiliary variables, expressing
the order between the grants in the system.

\section{Explanation of Counterexamples Using Causality}

Explanation of counterexamples addresses a  basic aspect of understanding a counterexample: the task of finding the failure in the trace itself. To motivate this approach, consider a verification engineer, who is formally verifying a hardware design written by a logic designer. The verification engineer writes a specification -- a temporal logic formula -- and runs a model checker, in order to check the formula on the design. If the formula fails on the design-under-test (DUT), a counterexample trace is produced and displayed in a trace viewer. The verification engineer does not attempt to debug the DUT implementation (since that is the responsibility of the the logic designer who wrote it). Her goal is to look for some basic information about the manner in which the formula fails on the specific trace. If the formula is a complex combination of several conditions, she needs to know which of these conditions has failed. These basic questions are prerequisites to deeper investigations of the failure. 
Ben-David et al. present a method and a tool for explaining the trace, without involving the model from which it was extracted \cite{BBCOT12}. This gives the approach the advantage of being light-weight,
as its running time depends only on the size of the specification and the counterexample, which is
much smaller than the size of the system. An additional advantage of the tool is that it is
independent on the verification procedure, and can be added as an external layer to any
tool, or even applied as an explanation of simulation traces. The main idea of the algorithm is to
represent the trace and the property that fails on this trace as a causal model and context
(see Section~\ref{cover-cause} on the description of the transformation). Then, the values of
signals in specific cycles are viewed as variables, and the set of causes for failure is marked
as red dots on the trace that is shown to the user graphically, in form of a timing diagram. The tool
is a part of the IBM RuleBase verification platform~\cite{RBurl} and is used extensively by the users.
While the trace is small compared to the system, the complexity of the exact algorithm for
computing causality leads to time-consuming computations; in order to keep the interactive nature
of the tool, the algorithm operates in one pass on the trace and 
computes an \emph{approximate} set of causes, which coincides with the
actual set of causes on all but contrived examples.
The reader is referred to \cite{BBCOT12} for more details of the implementation.

\section{Responsibility in Symbolic Trajectory Evaluation}

Symbolic Trajectory Evaluation (STE)~\cite{STEBryant} is a
powerful model checking technique for hardware verification, which
combines symbolic simulation with 3-valued abstraction. Consider a
circuit $M$, described as a Directed Acyclic Graph of nodes
that represent gates and latches. For such a circuit, an STE
assertion is of the form $A\rightarrow C$, where the
\emph{Antecedent A} imposes constraints over nodes of $M$ at
different times, and the \emph{Consequent C} imposes requirements
on $M$'s nodes at different times. 
The antecedent may introduce symbolic Boolean variables on some of the nodes.
The nodes that are not restricted by $A$ are initialized by
STE to the value $X$ ("unknown"), thus obtaining an $abstraction$ of the
checked model.

 STE is successfully used in the hardware industry for verifying
very large models with wide data
paths~\cite{forte,HighSchubert,CaseStudy}. The common method for
performing STE is by representing the values of each node in the
circuit by \emph{Binary Decision Diagrams (BDDs)}. 
To avoid the potential state explosion resulting from 
instantiating all unconstrained nodes, typically the circuit is refined
manually in iterations, until the value of the circuit is determined.

To avoid the need for manual refinement (which requires a close familiarity
with the structure of the circuit), \cite{CGY08} suggest to compute an approximation
of the \emph{degree of responsibility} of each node in the value of the output circuit.
Then, the instantiation can proceed in the order of decreasing degree of responsibility. The idea
behind this algorithm is that the nodes with the highest degree of responsibility are more likely
to influence the value of the circuit, and hence we will avoid instantiating too many nodes.
The algorithm was implemented in Intel STE framework for hardware verification 
and demonstrated better
results than manual refinement~\cite{CGY08}.

\section{\ldots and Beyond}

In formal verification, there is no natural application for the notion of blame (Def.~\ref{def-blame}),
since the model and the property are assumed to be known. On the other hand, in legal applications
it is quite natural to talk about an epistemic state of an agent, representing what the
agent knew or should have known. As~\cite{HP01b} points out, the legal system does not agree with
the structural definitions of causality, responsibility, and blame. However, it is still possible to
apply our methodology of representing a problem using causal models in order to improve our
understanding and analysis of particular situations. In \cite{CFKL15}, we make the case of
using the framework of actual causality in order to guide legal inquiry. In fact, the concepts of
responsibility and blame fit the procedure of legal inquiry very well, since we can capture both the
limited knowledge of the participants in the case, and the unknown factors in the case. We
use the case of baby P. -- a baby that died from continuous neglect and abuse, which was
completely missed by the social services and the doctor who examined him -- to demonstrate how
we can capture the known and unknown factors in the case and attempt to quantify the blame of
different parties involved in the case.

\bibliographystyle{eptcs}


\end{document}